# Experimental Research of the Diffraction and Vavilov-Cherenkov Radiation Generation in a Teflon Target

**M Shevelev, G Naumenko, A. Potylitsyn and Yu Popov**
Tomsk Polytechnic University, Lenina str. 2, Tomsk, 634050, Russia

E-mail: naumenko@tpu.ru

**Abstract.** Geometry of Vavilov-Cherekov (VChR) radiation when an electron moves close to a dielectric target is in analogy to diffraction radiation (DR) geometry. In this case we may expect DR generation from the upstream face of the target besides that VChR. The joint observation of these booth types of radiation is very interesting from the pseudo-photon viewpoint, which is applicable for relativistic electrons. Unexpected results obtained in our experiment insist on reflection about nature both DR and VChR. The experiment was performed on the relativistic electron beam of the microtron of Tomsk Polytechnic University.

## 1. About problem

The nature of Vavilov-Cherekov radiation (VChR) in dielectric targets, as well as the transition radiation, traditionally is considered as a dynamical polarization of the target material with macroscopic permittivity $\varepsilon$ by an electron electromagnetic field, followed by the propagation of this excitation in a target with velocity $v = c/n$, where $c$ is the speed of light in vacuum, $n$ is a refractive index [1]. In traditional representation, VChR appears when a particle moves in a medium with a velocity higher than the phase velocity of an electromagnetic wave in this medium. However, for high-energy electrons with the Lorenz factor $\gamma \gg 1$, the characteristic transverse size of an electron field $\gamma\lambda$ can be macroscopic (here, $\lambda = 2\pi c/\omega$ is the wavelength). Under these conditions, VChR can appear when the electron moves near the target [1-3] (see figure 1). Therewith the radiation angular distribution is satisfied to the well-known condition $\cos\theta_{ch} = 1/n\beta$, where $\beta$ is the electron velocity in units of $c$.

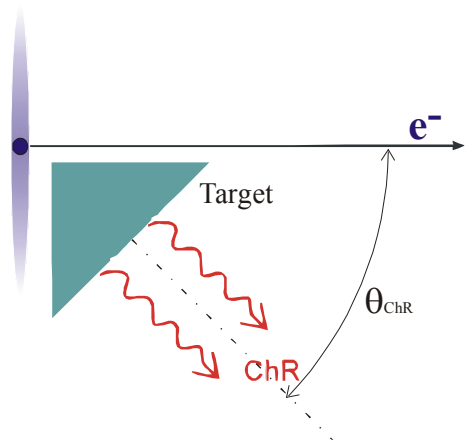

**Figure 1**. Geometry of VChR emission.

By analogy with transition radiation, the appearance of diffraction radiation (DR) should be expected due to the interaction of the field of the electron with the edge of the dielectric target; this radiation propagates inside the target in an angular cone that is determined by the velocity of the particle and the refractive index along the direction of electron motion (see figure 2).

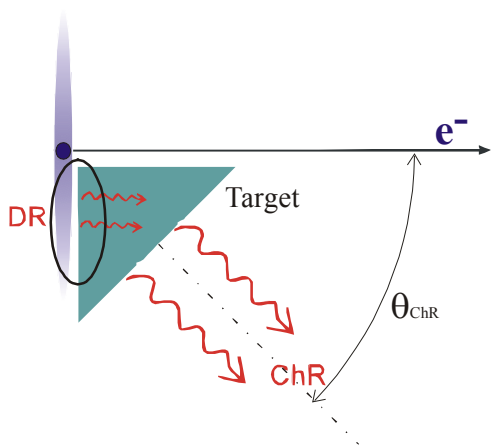

**Figure 2**. Geometry of VChR and DR simultaneously emission.

On the other hand, the pseudo-photons viewpoint may be also used for analyzes of an interaction of the relativistic electron electromagnetic field with matter. The "pseudo-photon" method proposed by Fermi [4] and developed by Williams [5] is widely used for theoretical studies of electromagnetic processes (see for instance [6] and [7]). According to this approach, the field of a charged particle may be replaced by a field of photons, which in this case are called pseudo-photons (in [8] is used the term "virtual quanta"; it should be differed on the same term in the quantum theory). This approach provides a good accuracy for the ultra-relativistic particles when the particle velocity is close to the light velocity ($v_p \approx c$) (see [6]) and when the longitudinal electric field component of the particle is negligible. In this case the particle field has the same properties as the field of real photons.

Using the pseudo-photons viewpoint, the DR in a dielectric target, when an electron moves close to the target, may be simply explained. When pseudo-photons of electron field cross the upstream face of a target, they continue to propagat in a target material like usual photons with the velocity $v = c/n$, taking into account the refraction on the upstream face. However, as for a large $n$ the pseudo-photons velocity is less then electron velocity, pseudo-photons lag behind the electron, and the electron becomes semi-bare.

There is the question, how can we explain the VChR using the pseudo-photons viewpoint? On the first view it seems impossible. This misunderstanding puts in doubt the possibility of application of this viewpoint. The experiment investigations presented in this paper are devoted to resolve this problem.

## 2. Experimental set-up and procedure

The experiment was carried out in the extracted electron beam of the Physical and Technical Institute of Tomsk Polytechnic University microtron with parameters presented in Table 1.

The experiment was performed on the extracted electron beam of the microtron. The beam is extracted from the vacuum chamber through a 50 $\mu m$ thick beryllium foil. The beam parameters are listed in Table 1.

**Table 1.** Electron beam parameters.

| | | | |
|---|---|---|---|
| Electron energy | 6.1 MeV ($\gamma = 12$) | Bunch period | 380 psec |
| Train duration | $\tau \approx 4$ $\mu$sec | Bunch population | $N_e = 6 \cdot 10^8$ |
| Bunches in a train | $n_b \approx 1.6 \cdot 10^4$ | Bunch length | $\sigma \approx 1.9 \sim 2.4$ mm |

For the radiation measurements the room temperature detector DP20M (figure 3) was used, with parameters described in [9]. Its main elements are a low-threshold diode, a broadband antenna and a preamplifier.

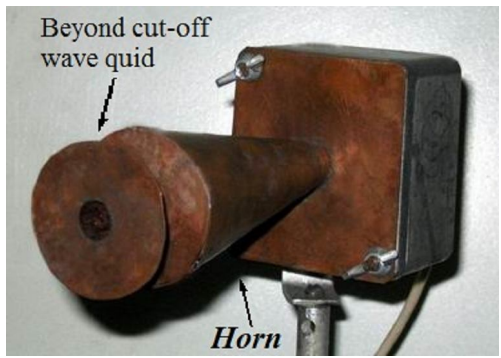

**Figure 3**. Detector DP20M.

The detector efficiency in the wavelength region $\lambda = 3 \sim 16$ mm is estimated to be constant to a $\pm 15\%$ accuracy. The detector sensitivity is 0.3 V/mWatt. A wave-guide with a cutoff $\lambda_{cut} = 17$ mm was used to cut the long-wave background of the accelerator RF system. The high frequency limit of the wavelength interval is defined by the bunch form-factor. This limit ($\lambda_{min} = 9$ mm) was measured using discrete wave filters [10] and a grating spectrometer. The coherent radiation intensity for $\lambda > 9$ mm) is by 8 orders larger than the incoherent one and has a power level =1 Watt per steradian. It means that one can investigate coherent radiation in this wavelength range without difficulty.

To exclude the prewave zone effect [11] a parabolic telescope (see figure 4) was used for the angular distribution measurement. This method was suggested and tested in [9] and gives the same angular distribution as in the far field zone ($R \gg \gamma^2 \lambda$).

The scheme of experimental set-up is shown in figure 4. Teflon target in the form of a triangular prism with dimensions $175 \times 175 \times 74$ mm was used in experiment. The target was placed at a distance of 220 mm from the extraction window; the distance from the target to the electron beam (impact parameter) was 20 mm. The refractive index for Teflon target was measured using the method described in [11] and was equal to n = 1.41 ± 0.1 in the wavelength range from 10 to 20 mm.

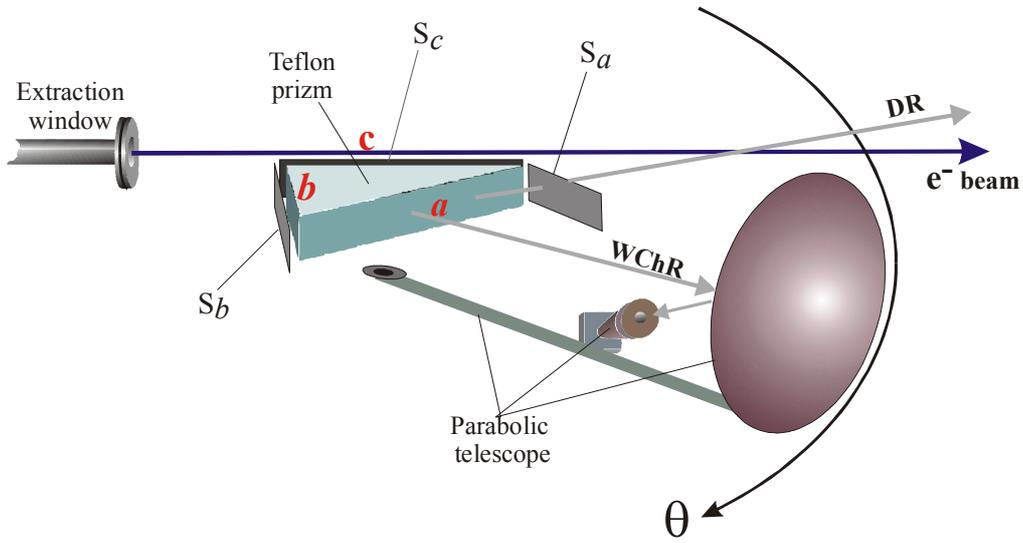

**Figure 4**. Scheme of VChR and DR simultaneously measurement. $S_a$, $S_b$ and $S_c$ are the removable conductive screens. *a*, *b*, *c* are the vertical faces of the target.

In experimental analysis, if necessary were used alternately conductive screens $S_a$, $S_b$ and $S_c$ (see Figure 4). Screens $S_b$ and $S_c$ were set close to the face of the prism.

## 3. Measurements

Measurements of the coherent radiation intensity on the angle θ were performed with a step $0.5°$. In figure 5 circles shows the measured radiation intensity on the angle without screens $S_a$, $S_b$ and $S_c$. The angle $\theta = 0°$ corresponds to the direction of the electron beam. The peak at the angle $\theta = 44°$ corresponds to VChR.

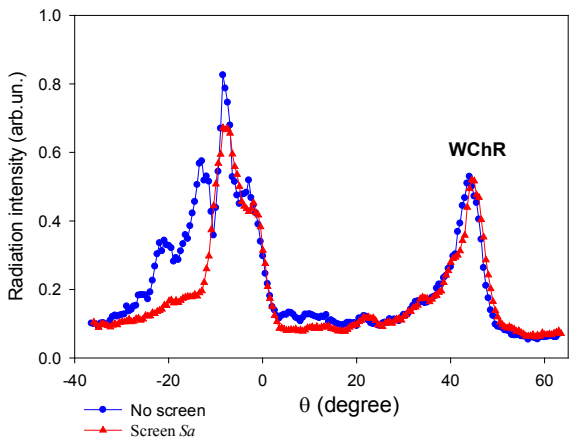
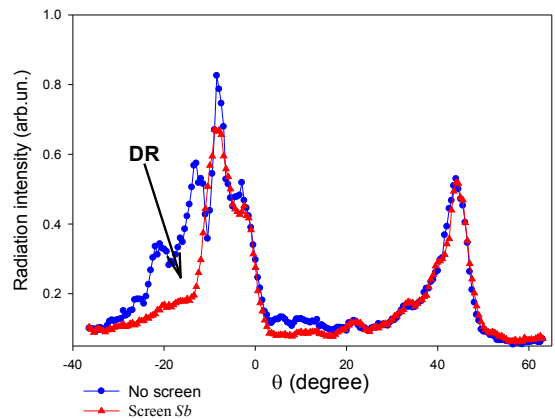

**Figure 5**. Measured angular distributions including screening of the face *a*.

**Figure 6**. Measured angular distributions including screening of the face *b*.

For the second step, the screen $S_a$ was installed. The radiation angular distribution for this geometry is shown in figure 5 by triangular points. It is clear, that the difference between these distributions (at the angle $\theta \approx -20°$) is a radiation from the downstream face *a* of the target. The peak at the angle $\theta = -8°$ corresponds to a backward transition radiation from the parabolic mirror.

To clarify the nature of the radiation at the angle $\theta \approx -20°$, the screen $S_a$ was removed and screen $S_b$ was installed. The radiation angular distribution for this geometry is shown in figure 6 by triangular points. The curve showed in this figure by circles, is the angular distribution without screens. One may see, that figures 5 and 6 are almost the same. Taking into account the shadowing effect from screen $S_b$ (see [13]), we can conclude, that the radiation at the angle $\theta \approx -20°$ is a DR from the upstream face $b$ of the target, which propagates inside the target and is refracted on the downstream face $a$.

One should be called attention to very important fact: the screen $S_b$ screens the DR but does not exert influence on the VChR (peak at angle $\theta = 44°$ in figure 6). Below we will turn back to this problem.

For more detail analyzes of the nature of radiation we had removed the screen $S_b$ and installed screen $S_c$ (see figure 4). The measured radiation angular distribution is shown in figure 7 (triangular points). For comparison the angular distribution without screens is shown by circles.

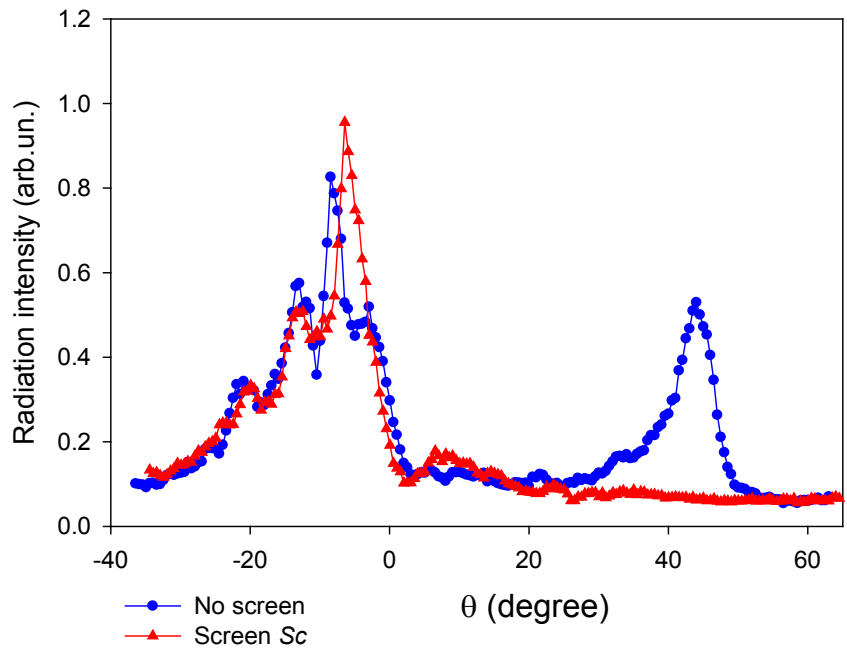

**Figure 7**. Measured angular distributions including screening of the face $c$.

One may see in figure 7, that in this geometry the VChR disappears. However the screen $S_c$ does not exert influence on the DR.

**4. Discussion**
Let us make the detail analysis of obtained experimental results. First of all the experiment results show the simultaneous generation of DR and VChR inside a target, and in doing so the DR is generated from the upstream face of the target, because the screen $S_c$ does not exert influence on the DR, but the screening of the upstream face $b$ (figure 4) suppresses a DR generation (according to [14] the surface currents are not induced on a downstream surface of a thick conductive screen, therefore radiation is not emitted from this screen in forward direction). This deduction is in agreement with both traditional and pseudo-photon viewpoints. It is clear that the electron field area up to radius about $\gamma\lambda$ affects on a DR generation.

Another generation mechanism we see for VChR. The screening by screen $S_b$ does not exert influence on the VChR. Ie VChR is not a result of bulk dielectric polarization. We can conclude from result, shown in figure 7, that VChR is generated from the face *c* of the prism, which is parallel to the electron beam trajectory. According to the pseudo-photons viewpoint, when pseudo-photons of electron field cross the upstream face of a target, they continue propagate in a target material like usual photons with the velocity $v = c/n$. For a large *n* the pseudo-photons velocity is less then electron velocity and pseudo-photons lag behind the electron, causing DR. In so doing, the electron becomes semi-bare and a shadow appears directly downstream to the face *b* (in figure 4) of the prism (see [13]). According to [15] this state is unstable and the electron field is recovered to the usual state at the distance about $\gamma^2 \lambda$. In the recovery process, the recovered electron field is refracted immediately inside the prism and also lags behind the electron. This process is continuous throughout the length of the target along the face *c*. In so doing, the refracted pseudo-photons become the sources of VChR for a direction, which is satisfied to the Vavilov-Cherenkov condition.

Finally we can conclude that macroscopic Maxwell equations may be used (in principle) for calculating the radiation emitted by relativistic electrons passing through or near material targets. Nevertheless phenomenological concepts like *pseudo-photons* and *semi-bare electron* sometimes are more useful for an intuitive understanding of the main features of radiation mechanisms. Our experimental analysis of radiation from a dielectric prism, when electrons move close to some face of the prism, shows that the VChR can be explained in frame of pseudo-photon viewpoint together with the *semi-bare electron recovery* concept of Feĭnberg [15]. In principle, as it was shown above, the *semi-bare electron* concept follows from the pseudo-photon viewpoint, but the interpretation of the electron field recovery process, needs the additional justification.

**Acknowledgment**
This work was partly supported by the warrant-order 1.226.08 of the Ministry of Education and Science of the Russian Federation and by the Grant RFBR 11-02-91177-GFEN_a.